\documentclass[english,aps,preprint]{revtex4}
\usepackage{pslatex}
\usepackage[T1]{fontenc}

\makeatletter

\usepackage{babel}
\makeatother
\begin{document}

\preprint{Aug 31, 2005, Rev Jan 2, 2006}

\title{Some Square Lattice Green Function Formulas}

\author{Stefan Hollos}

\email{stefan@exstrom.com}

\homepage{http://www.exstrom.com/stefan/stefan.html}

\affiliation{Exstrom Laboratories LLC, 662 Nelson Park Dr, Longmont, Colorado
80503, USA.}

\author{Richard Hollos}

\affiliation{Exstrom Laboratories LLC, 662 Nelson Park Dr, Longmont, Colorado
80503, USA.}

\begin{abstract}
We derive formulas for the matrix elements of the two dimensional
square lattice Green function along the diagonal, and along the coordinate
axes. We also give an asymptotic formula for the diagonal elements.
\end{abstract}
\maketitle

\section{introduction}

In some recent papers \cite{cserti00,cserti02} Cserti showed how
a lattice Green function (LGF) can be used to find the resistance
between two points in an infinite lattice of resistors. Cserti gives
an expression for the matrix elements of the LGF in the form of an
integral. In this paper we will show how to solve this integral for
the case of a two dimensional square lattice along the diagonal and
the coordinate axes. This allows any arbitrary diagonal or coordinate
axis LGF matrix element to be calculated directly. Formulas for these
elements were first derived by McCrea and Whipple \cite{comment}\cite{mccrea40}
using a different procedure than that presented here. We will also
give an asymptotic formula for the diagonal matrix elements that converges
to Cserti's asymptotic limit formula for large values of $n$.

The LGF with which we are concerned here can in general be used to
solve the discrete two dimensional Poisson equation with boundary
conditions at infinity. Therefore it will be useful in solving two
dimensional electrostatics problems \cite{exstrom2005} as well as
many other problems that can be modeled by a Poisson equation.

\section{diagonal matrix elements}

The matrix elements of the two dimensional square lattice Green function
can be expressed in terms of an integral as\begin{equation}
g(n,m)=\frac{1}{2\pi^{2}}\int_{0}^{\pi}d\phi\int_{0}^{\pi}d\theta\frac{1-\cos n\theta\cos m\phi}{2-\cos\theta-\cos\phi}\label{eq:1}\end{equation}
This is essentially the same as Cserti's \cite{cserti00} eq. B1.
We will begin by looking at the diagonal elements where $m=n$. First
note that for $m=n$ eq. \ref{eq:1} can be rewritten in the following
form\begin{equation}
g(n,n)=\frac{1}{2\pi^{2}}\int_{0}^{\pi}d\phi\int_{0}^{\pi}d\theta\frac{1-\frac{1}{2}\cos n(\phi-\theta)-\frac{1}{2}\cos n(\phi+\theta)}{2-2\cos\frac{1}{2}(\phi-\theta)\cos\frac{1}{2}(\phi+\theta)}\label{eq:2}\end{equation}
By symmetry the two cosine terms in the numerator of the integrand
can be combined to give\begin{equation}
g(n,n)=\frac{1}{2\pi^{2}}\int_{0}^{\pi}d\phi\int_{0}^{\pi}d\theta\frac{1-\cos n(\phi+\theta)}{2-2\cos\frac{1}{2}(\phi-\theta)\cos\frac{1}{2}(\phi+\theta)}\label{eq:3}\end{equation}
In terms of new variables $\phi'=\frac{1}{2}(\phi-\theta)$, $\theta'=\frac{1}{2}(\phi+\theta)$
eq. \ref{eq:3} becomes\begin{equation}
g(n,n)=\frac{1}{4\pi^{2}}\int_{0}^{\pi}d\phi\int_{0}^{\pi}d\theta\frac{1-\cos2n\theta}{1-\cos\phi\,\cos\theta}\label{eq:4}\end{equation}
The integration over $\phi$ can now be done to give\begin{equation}
g(n,n)=\frac{1}{4\pi}\int_{0}^{\pi}d\theta\,\frac{1-\cos2n\theta}{\sin\theta}\label{eq:5}\end{equation}
Using the identity $1-\cos2n\theta=2\sin^{2}n\theta$, eq. \ref{eq:5}
can be written as\begin{equation}
g(n,n)=\frac{1}{2\pi}\int_{0}^{\pi}d\theta\,\frac{\sin^{2}n\theta}{\sin\theta}=\frac{1}{2\pi}\int_{0}^{\pi}d\theta\,\left(\frac{\sin n\theta}{\sin\theta}\right)^{2}\sin\theta\label{eq:6}\end{equation}
Now if we let $x=\cos\theta$ then this becomes the integral of a
type II Chebyshev polynomial \cite{mason2003}\begin{equation}
g(n,n)=\frac{1}{2\pi}\int_{-1}^{1}U_{n-1}^{2}(x)\, dx=\frac{1}{\pi}\int_{0}^{1}U_{n-1}^{2}(x)\, dx\label{eq:7}\end{equation}
The square of a type II Chebyshev polynomial can be expressed as\begin{equation}
U_{n}^{2}(x)=\sum_{k=0}^{n}U_{2k}(x)\label{eq:8}\end{equation}
To prove this identity it is sufficient to show that $U_{n}^{2}(x)-U_{n-1}^{2}(x)=U_{2n}(x)$.
With $x=\cos\theta$ we have $U_{n}(x)=\sin(n+1)\theta/\sin\theta$
and \[
U_{n}(x)-U_{n-1}(x)=\frac{\sin(n+1)\theta-\sin n\theta}{\sin\theta}=\frac{\cos(n+\frac{1}{2})\theta}{\cos\frac{1}{2}\theta}\]
\[
U_{n}(x)+U_{n-1}(x)=\frac{\sin(n+1)\theta+\sin n\theta}{\sin\theta}=\frac{\sin(n+\frac{1}{2})\theta}{\sin\frac{1}{2}\theta}\]
\[
U_{n}^{2}(x)-U_{n-1}^{2}(x)=\frac{\cos(n+\frac{1}{2})\theta\;\sin(n+\frac{1}{2})\theta}{\cos\frac{1}{2}\theta\;\sin\frac{1}{2}\theta}=\frac{\sin(2n+1)\theta}{\sin\theta}=U_{2n}(x)\]

So using eq. \ref{eq:8}, eq. \ref{eq:7} can be written as\begin{equation}
g(n,n)=\frac{1}{\pi}\sum_{k=0}^{n-1}\int_{0}^{1}U_{2k}(x)\, dx\label{eq:9}\end{equation}
Letting $x=\cos\theta$, the integrals in eq. \ref{eq:9} become\begin{equation}
\int_{0}^{1}U_{2k}(x)\, dx=\int_{0}^{\frac{\pi}{2}}\sin(2k+1)\theta\, d\theta=\frac{1}{2k+1}\label{eq:10}\end{equation}
Substituting this into eq. \ref{eq:9} and changing the summation
index gives\begin{equation}
g(n,n)=\frac{1}{\pi}\sum_{k=1}^{n}\frac{1}{2k-1}\label{eq:11}\end{equation}
Note that $g(n,n)$ also obeys a difference equation and that the
solution for $g(n,n)$ given in eq. \ref{eq:11} could also be arrived
at by solving the difference equation. The difference equation for
$g(n,n)$ is the same as that given by Cserti \cite{cserti00} eq.
32 for the resistances along the diagonal\begin{equation}
(2n+1)\, g(n+1,n+1)-4n\, g(n,n)+(2n-1)\, g(n-1,n-1)=0\label{eq:11A}\end{equation}
Since the coefficients of this equation add up to zero, if we substitute
$g(n,n)=\sum_{k=1}^{n}f(k)$, $g(n-1,n-1)=g(n,n)-f(n)$, $g(n+1,n+1)=g(n,n)+f(n+1)$
into the equation, we will get a first order equation for $f(n)$.\begin{equation}
(2n+1)\, f(n+1)-(2n-1)\, f(n)=0\label{eq:11B}\end{equation}
With the initial condition $f(1)=\frac{1}{\pi}$ the solution to this
equation is $f(k)=\frac{1}{\pi}\,\frac{1}{2k-1}$ which once again
gives us eq. \ref{eq:11} as the solution for $g(n,n)$.

We will now derive an asymptotic formula for $g(n,n)$. First note
that the $g(n,n)$ elements are proportional to the partial sums of
a generalized harmonic series. They can also be expressed in terms
of the standard harmonic series as follows\begin{equation}
g(n,n)=\frac{1}{\pi}\left(\sum_{k=1}^{2n}\frac{1}{k}-\frac{1}{2}\sum_{k=1}^{n}\frac{1}{k}\right)=\frac{1}{\pi}\left(H_{2n}-\frac{1}{2}H_{n}\right)\label{eq:12}\end{equation}
where we have introduced the notation\begin{equation}
H_{n}=\sum_{k=1}^{n}\frac{1}{k}\label{eq:13}\end{equation}
for the $n$th partial sum of the standard harmonic series. The asymptotic
formula for the $n$th partial sum of the harmonic series is \cite{arfken1985}
p. 338\begin{equation}
H_{n}=\ln n+\gamma+\frac{1}{2n}-\sum_{k=1}\frac{B_{2k}}{2kn^{2k}}\label{eq:13A}\end{equation}
where $\gamma=0.5772156649\ldots$ is the Euler-Mascheroni constant
and $B_{2k}$ is a Bernoulli number. Using this in eq. \ref{eq:12}
results in the following asymptotic formula for $g(n,n)$

\begin{equation}
g(n,n)=\frac{1}{2\pi}\left[\ln(n)+\gamma+2\ln(2)+\sum_{k=1}^{\infty}\frac{B_{2k}(2^{2k-1}-1)}{k(2n)^{2k}}\right]\label{eq:14}\end{equation}
Without the Bernoulli sum, this is essentially the same as Cserti's
\cite{cserti00} eq. 33 for the asymptotic limit of the resistance.

\section{the on-axis elements}

We now turn to the on-axis elements where $m=0$ and eq. \ref{eq:1}
becomes\begin{equation}
g(n,0)=\frac{1}{2\pi^{2}}\int_{0}^{\pi}d\phi\int_{0}^{\pi}d\theta\frac{1-\cos n\theta}{2-\cos\theta-\cos\phi}\label{eq:15}\end{equation}
The integral with respect to $\phi$ can be carried out to give\begin{equation}
g(n,0)=\frac{1}{2\pi}\int_{0}^{\pi}d\theta\frac{1-\cos n\theta}{\sqrt{(2-\cos\theta)^{2}-1}}\label{eq:16}\end{equation}
Now note that $1-\cos n\theta=2\sin^{2}(\frac{n\theta}{2})$ and $2-\cos\theta=1+2\sin^{2}(\frac{\theta}{2})$
so that the denominator of the integrand in eq. \ref{eq:16} becomes
$\sqrt{(1+2\sin^{2}(\frac{\theta}{2}))^{2}-1}=2\sin(\frac{\theta}{2})\sqrt{1+\sin^{2}(\frac{\theta}{2})}$.
Eq. \ref{eq:16} then becomes\begin{equation}
g(n,0)=\frac{1}{2\pi}\int_{0}^{\pi}d\theta\frac{\sin^{2}(\frac{n\theta}{2})}{\sin(\frac{\theta}{2})\sqrt{1+\sin^{2}(\frac{\theta}{2})}}\label{eq:18}\end{equation}
Making the change in variable $\theta'=\frac{\theta}{2}$ , we write
eq. \ref{eq:18} as\begin{equation}
g(n,0)=\frac{1}{\pi}\int_{0}^{\frac{\pi}{2}}d\theta\left(\frac{\sin n\theta}{\sin\theta}\right)^{2}\frac{\sin\theta}{\sqrt{1+\sin^{2}\theta}}\label{eq:19}\end{equation}
With $x=\cos\theta$ we then have an integral involving the Chebyshev
polynomial $U_{n-1}(x)$\begin{equation}
g(n,0)=\frac{1}{\pi}\int_{0}^{1}\frac{U_{n-1}^{2}(x)}{\sqrt{2-x^{2}}}\, dx\label{eq:20}\end{equation}
Now if we let $x=\sqrt{1-\cos\theta}$ then\begin{equation}
g(n,0)=\frac{1}{2\pi}\int_{0}^{\frac{\pi}{2}}U_{n-1}^{2}(\sqrt{1-\cos\theta})\, d\theta\label{eq:21}\end{equation}
\begin{equation}
g(n,0)=\frac{1}{2\pi}\sum_{k=0}^{n-1}\int_{0}^{\frac{\pi}{2}}U_{2k}(\sqrt{1-\cos\theta})\, d\theta\label{eq:22}\end{equation}
If the type I and type II Chebyshev polynomials are expressed in the
following forms \cite{mason2003}\begin{eqnarray}
T_{n}(x) & = & \frac{1}{2}\left[\left(x+\sqrt{x^{2}-1}\right)^{n}+\left(x-\sqrt{x^{2}-1}\right)^{n}\right]\label{eq:23}\\
U_{n}(x) & = & \frac{\left(x+\sqrt{x^{2}-1}\right)^{n+1}-\left(x-\sqrt{x^{2}-1}\right)^{n+1}}{2\sqrt{x^{2}-1}}\nonumber \end{eqnarray}
then it is straightforward to prove the identity\begin{equation}
U_{2k}(\sqrt{1-x})=\frac{(-1)^{k}T_{2k+1}(\sqrt{x})}{\sqrt{x}}\label{eq:24}\end{equation}
Using this identity eq. \ref{eq:22} becomes\begin{equation}
g(n,0)=\frac{1}{2\pi}\sum_{k=0}^{n-1}(-1)^{k}\int_{0}^{\frac{\pi}{2}}\frac{T_{2k+1}(\sqrt{\cos\theta})}{\sqrt{\cos\theta}}\, d\theta\label{eq:25}\end{equation}
$T_{2k+1}(x)$ can be expressed in series form as\begin{equation}
T_{2k+1}(x)=\sum_{j=0}^{k}(-1)^{k-j}2^{2j}\frac{2k+1}{2j+1}\left(\begin{array}{c}
k+j\\
2j\end{array}\right)x^{2j+1}\label{eq:26}\end{equation}
so that we have\begin{equation}
(-1)^{k}\frac{T_{2k+1}(\sqrt{x})}{\sqrt{x}}=\sum_{j=0}^{k}(-4)^{j}\frac{2k+1}{2j+1}\left(\begin{array}{c}
k+j\\
2j\end{array}\right)x^{j}\label{eq:27}\end{equation}
and eq. \ref{eq:25} can be written as\begin{equation}
g(n,0)=\frac{1}{2\pi}\sum_{k=0}^{n-1}\,\sum_{j=0}^{k}a(k,j)b(j)\label{eq:28}\end{equation}
where\begin{eqnarray}
a(k,j) & = & (-4)^{j}\frac{2k+1}{2j+1}\left(\begin{array}{c}
k+j\\
2j\end{array}\right)\label{eq:29}\\
b(j) & = & \int_{0}^{\frac{\pi}{2}}\cos^{j}\theta\, d\theta=\left\{ \begin{array}{cc}
\frac{\pi}{2}\frac{(j-1)!!}{j!!} & j=0,2,4,6,\ldots\\
\frac{(j-1)!!}{j!!} & j=1,3,5,7,\ldots\end{array}\right.\nonumber \end{eqnarray}
Eq. \ref{eq:28} can be used to directly calculate $g(n,0)$ for arbitrary
values of $n$.

\section{conclusion}

We have derived equations by which $g(n,n)$ and $g(n,0)$ can be
calculated for arbitrary values of $n$. For the case of $g(n,n)$
we have an asymptotic formula eq. \ref{eq:14} that allows for a quick
and efficient calculation. In the case of $g(n,0)$ we have eq. \ref{eq:28}
whose evaluation can be optimized for large values of $n$. A complete
asymptotic formula for $g(n,0)$ has not yet been found. A formula
very similar to eq. \ref{eq:14} for $g(n,n)$ has been found for
$g(n,0)$ but only the first few terms in the Bernoulli sum have so
far been determined. Formulas for the general matrix elements $g(n,m)$
have been found by us. These formulas are found by solving the partial
difference equation for $g(n,n)$. This equation can only be solved
after a formula for the diagonal elements, $g(n,n)$ has been found.

\begin{acknowledgments}
The authors acknowledge the generous support of Exstrom Laboratories
and its president Istvan Hollos.

\bibliographystyle{apsrev}
\bibliography{../lgfint/gf,lgfmath1}

\end{acknowledgments}

\end{document}